\documentstyle[12pt,fleqn,aaspp4]{article}
\newcommand{\be}{\begin{equation}}
\newcommand{\ee}{\end{equation}}

\def\au{{\rm AU}}

\def\yr{{\rm yr}}

\def\dol{{d_{\rm ol}}}
\def\dls{{d_{\rm ls}}}
\def\dos{{d_{\rm os}}}

\def\day{{\rm day}}

\def\kms{{\rm km}\,{\rm s}^{-1}}
\def\bv{{\bf v}}
\def\bu{{\bf u}}

\begin{document}

\title{Microlens Parallaxes with {\it SIRTF}}

\author{Andrew Gould\altaffilmark{1}\altaffiltext{1}
{Alfred P.\ Sloan Foundation Fellow}}
\affil{Ohio State University, Department of Astronomy, 
174 West 18th Ave., Columbus, OH 43210}
\affil{E-mail: gould@astronomy.ohio-state.edu}

\begin{abstract}

	The {\it Space Infrared Telescope Facility (SIRTF)} will drift away 
from the Earth at $\sim 0.1\,\au\,\yr^{-1}$.  Microlensing events will 
therefore have different characteristics as seen from the satellite and the 
Earth.  From the difference, it is possible in principle to measure
$\tilde \bv$, the transverse
velocity of the lens projected onto the observer plane.
Since $\tilde \bv$ has very different values for different populations
(disk, halo, Large Magellanic Cloud), such measurements could help identify 
the location, and hence the nature, of the lenses.  I show that the method 
previously developed by Gould for measuring such satellite parallaxes fails 
completely in the case of {\it SIRTF}: it is overwhelmed by degeneracies which 
arise from fact that the Earth and satellite observations are in different 
band passes.  I develop a new method which allows for observations in 
different band passes and yet removes all degeneracies.  The method combines
a purely ground-based measurement of the ``parallax asymmetry'' with a 
measurement of the delay between the time the event peaks at the Earth
and satellite.  In effect, the parallax asymmetry determines the component
of $\tilde \bv$ in the Earth-Sun direction, while the delay time measures
the component of $\tilde \bv$ in the direction of the Earth's orbit.

\end{abstract}

\section{Introduction}

	Over a dozen candidate microlensing events have been discovered toward
the Large Magellanic Cloud (LMC) (Aubourg et al.\ 1993; Alcock et al.\ 1997a).
They have typical Einstein crossing times $t_e\sim 40\,$days.  Here, $t_e$
is related to the Einstein radius, $r_e$ by
\begin{equation}
t_e \equiv {r_e\over |\bv|},\quad r_e^2= {4 G M\dol\dls\over \dos c^2},
\label{eqn:tedef}
\end{equation}
where $\bv$ is the transverse
velocity of the lens relative to the observer-source line
of sight, $M$ is the mass, and $\dol$, $\dls$, and $\dos$ are the distances
between the observer, lens, and source.
If the lenses are assumed to be in the Galactic halo, they would appear to
make up of order half the dark matter and have 
typical masses $M\sim 0.4\,M_\odot$
(Alcock et al.\ 1997a).  Several lines of reasoning suggest that this 
interpretation is implausible.  However, to date there are no plausible
alternatives.

	The halo cannot be composed of $M\sim 0.4\,M_\odot$ hydrogen objects
or they would burn and would easily be detected.  If it were composed of
white dwarfs, the white dwarfs themselves, their progenitors at high redshift,
and the metals these produce would all be detectable in various ways
(Fields, Freese, \& Graff 1998).  The most viable candidates for halo lenses
seem to be exotic new objects such as primordial black holes -- which just
happen to have the same masses as the most common stars.  In addition, if
halo objects are causing events toward the LMC, they should also generate
events of similar duration and frequency toward the Small Magellanic Cloud
(SMC).  However, the only two events discovered toward the SMC to date
show significant evidence of being due to SMC lenses 
(Palanque-Delabrouille et al.\ 1998; Afonso et al.\ 1998; Alcock et al.\ 1997c,
1998; Albrow et al.\ 1998).

	On the other hand, if the LMC lensing events were due to lenses in the 
LMC bar/disk itself (Sahu 1994; Wu 1994), then the event 
rate should be much lower than observed (Gould 1995b), and their distribution 
on the sky should be more concentrated toward the bar (Alcock et al.\ 1997a).
If they were due to a tidally disrupted dwarf galaxy along the line of sight
to the LMC (Zhao 1997; Zaritsky \& Lin 1998), they should be observable
in surface brightness maps (Gould 1998a) and tracer populations like RR Lyrae
stars (Alcock et al.\ 1997b) and clump giants (Bennett 1998).  Hence, there
are no compelling candidates for the lens population.

	The different possible lens populations have radically different
kinematics, and one could therefore distinguish among them if kinematic
parameters could be measured.  This is not possible for most events because
the one measured quantity, $t_e$ is a combination of the mass, distance, and
speed (see eq.\ \ref{eqn:tedef}).  However, 
if a satellite were launched into solar orbit, the event
would appear different from the satellite
than from the Earth.  From the difference, one
could in effect measure the length of time it takes for the projected
position of the lens to travel from the Earth to the satellite and also
its direction of transverse motion.  Since the distance between the Earth
and satellite is known, one could thereby determine the two components of the
transverse velocity projected onto the plane of the observer,
\begin{equation}
\tilde \bv = {\dos\over\dls}\bv.\label{eqn:vtildedef}
\end{equation}
Since $\tilde v\equiv |\tilde \bv|$ has values of $\sim 50\,\kms$ for
disk lenses, $\sim 275\,\kms$ for halo lenses, and $\ga 1000\,\kms$ for
LMC lenses, measurement of this quantity should distinguish well among
components.

\section{Degeneracies}

	Within the framework originally formulated by Refsdal (1966) and 
Gould (1994), the parallax-satellite measurement of $\tilde \bv$ was subject to
a four-fold degeneracy, including a two-fold degeneracy in $\tilde v$.  These
degeneracies
can be understood as follows.  For any given observer (Earth or satellite),
the magnification $A(u)$ is a function only of the projected lens-source 
separation $u$ (Paczy\'nski 1986)
\begin{equation}
A(u) = {u^2 + 2\over u(u^2+4)^{1/2}},\qquad
u(t)= \biggl[\biggr({t-t_0\over t_e}\biggr)^2 + \beta^2\biggr]^{1/2},
\label{eqn:adef}
\end{equation}
where $t_0$ is the time of maximum magnification and $\beta$ is the impact
parameter.  In units
of the projected Einstein ring $\tilde r_e = (\dos/\dls)r_e$, the separation 
between the Earth and the satellite along the 
direction of the lens motion is simply $\Delta t_0/t_e$, where
\begin{equation}
\Delta t_0 \equiv t_{0,S} - t_{0,\oplus}\label{eqn:deltatnought}
\end{equation}
is the difference in the observed times of maximum.  However, the separation in
the direction normal to the lens motion (in units of $\tilde r_e$) has four
possible values, $\pm \Delta\beta_+$ and $\pm \Delta\beta_-$, where
$\Delta \beta_\pm=|\beta_S \pm \beta_\oplus|$.  Since
\begin{equation}
\tilde \bv\cdot \Delta \bu = {d \over t_e},\qquad
\Delta \bu \equiv (\Delta t_0/t_e,\Delta\beta),
\label{eqn:tildev}
\end{equation}
this engenders a four-fold degeneracy in $\tilde \bv$
in direction and a two-fold degeneracy
in amplitude.  Here $d$ is the Earth-satellite distance.

	It is, however, possible to break this degeneracy by taking advantage
of the fact that the velocities of the Earth and satellite differ in the
$\Delta \beta$ direction, leading to a difference in observed time scales,
$\Delta t_e = 2\pi\yr^{-1}\Delta \beta t_e^2$, where 
$\Delta t_e \equiv t_{e,S}-t_{e,\oplus}$, and where I have adopted the
conventions that the satellite is trailing the Earth and that if $\tilde\bv$
is outward from the Sun, then $\Delta\beta>0$.
  Hence, if this time difference can
be measured, $\Delta\beta$ can be unambiguously determined (Gould 1995a),
\begin{equation}
\Delta \beta = {1\over \Omega_\oplus t_e}\,{\Delta t_e\over t_e},\qquad
\Omega_\oplus \equiv {2\pi\over \yr}\sim (58\,\day)^{-1}.\label{eqn:betares}
\end{equation}

	However, measurement of the time-scale difference introduces an
additional degeneracy which must be dealt with very carefully.  The
observed flux from the star is actually a function of five parameters
\begin{equation}
F(t;t_0,\beta,t_e,F_0,B) = F_0 A[u(t;t_0,\beta,t_e)] + B,
\label{eqn:foft}
\end{equation}
where $F_0$ is the source flux and $B$ is any background light that falls
into the aperture but is not lensed.  Measurement of $B$ is highly correlated
both with $\beta$ and $t_e$ because the effect on the light curve of with
changing all of these three parameters is even in $(t-t_0)$ and very similar
to one another. Hence, it is all but impossible to measure
the small time-scale difference between the Earth and the satellite if
each light curve is fitted with its own background parameter, $B$.  When I
proposed this method (Gould 1995a), I 
therefore explicitly assumed that the background light was identical
for the two sources, $B_S \equiv B_\oplus$.
Physically, this condition can be
attained if the filters have identical (or nearly identical) transmission
properties, and if the images are convolved to the same seeing.  With this
constraint, $t_{e,S}$ is still highly correlated with $B_S$, and 
$t_{e,\oplus}$ is highly correlated with 
$B_\oplus$.  However, since $B_S = B_\oplus$, the 
difference $\Delta t_e\equiv t_{e,S}-t_{e,\oplus}$ is well constrained.

	The {\it Space Infrared Telescope Facility (SIRTF)} will be launched
into solar orbit early in the next decade, and will drift away from the
Earth at $\sim 0.1\,\au\,\yr^{-1}$.  In this respect, it is in an excellent
position to measure microlens parallaxes.  However, the bluest band that
it can observe is $L$ band ($\sim 3.6\,\mu$m).  Because of the complexity of
atmospheric transmission at $L$, it is not possible to mimic the space-based
detector response from the ground.  Moreover, the ground-based background in 
$L$
is so high that it is not practical to monitor typical LMC sources $(V\ga 20)$
from the ground.  Hence, one cannot insure $B_S=B_\oplus$, and so it is not 
possible for {\it SIRTF} to measure parallaxes using my original 
single-satellite approach (Gould 1995a).

\section{A New Parallax Method}

	To recapitulate, {\it SIRTF} has no trouble measuring one component
of $\Delta \bu$, namely $\Delta t_0$, which is unambiguously given by the
difference in times of peak magnifications.  However, it cannot measure the 
other component, $\Delta \beta$, either directly or through measurement
of $\Delta t_e$ (via eq.\ \ref{eqn:betares}) because both parameters are
degenerate with the blended light, $B$.  The key to resolving this problem
is to notice that equation (\ref{eqn:betares}) arises from the Earth-satellite
velocity difference in the $\Delta \beta$ direction.  However, the Earth
itself undergoes a velocity change in this direction during the course
of the event by an amount $\Delta v\sim \Omega t_e v_\oplus$, where
$v_\oplus=30\,\kms$ is the speed of the Earth.  This purely ground-based
parallax effect leads to an asymmetry in the light curve which can, in effect,
be used to measure $\Delta \beta$ and hence complete the measurement of
$\Delta \bu$.

	Actually, such parallax asymmetries have a long history.  I showed
that for events of sufficiently long duration, it would be possible to measure
a complete parallax from the ground (Gould 1992).  One such parallax 
measurement has been published for a long event ($t_e\sim 110\,$days)
seen toward the bulge
(Alcock et al.\ 1995) and several others have been observed (Bennett 1997).
For shorter events ($\Omega_\oplus t_e\la 1$), the acceleration of the Earth 
can
be approximated as constant over the course of the event.  Equation
(\ref{eqn:adef}) is then replaced by 
(Gould, Miralda-Escud\'e, \& Bahcall 1994)
\begin{equation}
u(t;t_0,\beta,t_e,\gamma) = 
\biggl[\biggl[\xi\biggl({t-t_0\over t_e}\biggr)\biggr]^2+
\beta^2\biggr]^{1/2},\qquad  \xi(y) 
= y + {1\over 2}\gamma y^2,\label{eqn:xpoft}
\end{equation}
where
\begin{equation}
\gamma = {v_\oplus\over \tilde v}\,\Omega_\oplus t_e\cos\phi,
\label{eqn:gammadef}
\end{equation}
$\phi$ is the angle between $\bv$ and the Earth-Sun separation,
and where I have made use of the fact that the LMC is approximately at the 
ecliptic pole.  Gould et al.\ (1994) noted that even for very short events,
halo lenses ($\tilde v\sim 275\,\kms$) could be distinguished from disk lenses
($\tilde v\sim 50\,\kms$), at least statistically.  The problem is that 
in any individual case, if
$\gamma$ were measured to be consistent with zero, one would not know
whether the lens were in the halo (where $\tilde v$ is large) or in the disk 
(and $\cos\phi$
just happened to be small).  Moreover, the principal question about the
location of the lenses is not halo vs.\ disk, but halo vs.\ LMC.  For short
events it would be extremely difficult to distinguish between  halo and LMC
lenses using this technique, even statistically.

	However, the typical events observed toward the LMC now turn out
to be considerably longer, $t_e\sim 40\,$days.  Indeed, their long time
scale is a major puzzle.  Hence, I recently proposed that LMC events be
intensively monitored to search for this effect (Gould 1998b).  I showed
that one could distinguish statistically between the halo-lens and
LMC-lens scenarios by observing 15--30 events over 5 years.
The GMAN collaboration (Alcock et al.\ 1997d)
is routinely monitoring LMC events, but probably not
intensively enough to detect this effect.  In any event, the fact that
one measures
only the degenerate combination, $\tilde v\sec\phi$, and not the two parameters
separately, means that one cannot resolve the nature of the events on an
individual basis.

\subsection{Overview}

	The new approach is to combine a ground-based measurement of $\gamma$
(and of course $t_{0,\oplus}$) with a space-based measurement of $t_{0,S}$
(and therefore $\Delta t_0$) to completely determine $\Delta \bu$.  With
this is mind, I write $\bu$ in a basis that is rotated relative to
equation (\ref{eqn:tildev}),
\begin{equation}
\Delta \bu = (\Delta u_x,\Delta u_y),\quad
\Delta u_x = {\Delta t_0\over t_e}\cos\theta + \Delta \beta\sin\theta,\quad
\Delta u_y = -{\Delta t_0\over t_e}\sin\theta + \Delta \beta\cos\theta
\label{eqn:budef}
\end{equation}
Here, $\Delta u_x$  and $\Delta u_y$ 
are anti-parallel and outward normal to the direction of the 
Earth's motion at the midpoint of the event, while $\Delta t_0$ and
$\Delta\beta$ are anti-parallel and normal to the satellite-Earth separation
vector.  The rotation angle $\theta$ is defined by
\begin{equation}
\sin{\theta}\equiv {d\over 2\au}.
\label{eqn:deltaux}
\end{equation}
This is convenient because $\Delta u_y$
can be simply expressed in terms of observables (see eqs.\
\ref{eqn:tildev} and \ref{eqn:gammadef}),
\begin{equation}
\Delta u_y =
{d\over\au}\,(\Omega_\oplus t_{e,\oplus})^{-2}\gamma_\oplus.
\label{eqn:deltauy}
\end{equation}

From both a conceptual and a practical point of view, it is useful to think
of the measurement process as first determining the four Earth parameters,
$t_{0,\oplus}$, $\beta_\oplus$, $t_{e,\oplus}$, and $\gamma_\oplus$, and
then using these to predict the four analogous satellite quantities
as a function of the (unknown) parameter $\Delta u_x$.  
\begin{equation}
t_{0,S} = t_{0,\oplus} + \Delta t_0,\qquad
{\Delta t_0\over t_{e,\oplus}} = \Delta u_x \cos\theta - 
2(\Omega_\oplus t_{e,\oplus})^{-2}\gamma_\oplus
\sin^2\theta
\label{eqn:sattnought}
\end{equation}
\begin{equation}
\beta_S = |\beta_\oplus \pm \Delta \beta|,\qquad
\Delta \beta = \Delta u_x \sin\theta +
(\Omega_\oplus t_{e,\oplus})^{-2}\gamma_\oplus
\sin 2\theta 
\label{eqn:satbeta}
\end{equation}
\begin{equation}
t_{e,S} = t_{e,\oplus} + \Delta t_e,\qquad
{\Delta t_e\over t_{e,\oplus}} = \Delta u_x \Omega_\oplus t_{e,\oplus}
\sin\theta + 
(\Omega_\oplus t_{e,\oplus})^{-1}\,\gamma_\oplus\sin 2\theta
\label{eqn:satte}
\end{equation}
\begin{equation}
\gamma_S = \Delta u_x (\Omega_\oplus t_{e,\oplus})^{2}
\cos\theta +
\gamma_\oplus \cos 2\theta 
\label{eqn:satgamma}
\end{equation}
The satellite
measurements are then used to determine $\Delta u_x$.

	Equations (\ref{eqn:satbeta}) and (\ref{eqn:satte}) yield very
little information about $\Delta u_x$ because
$\beta$ and $t_e$ are poorly determined, being strongly 
correlated with $B$.  
In addition, equation (\ref{eqn:satbeta}) is ambiguous because, in 
the limit $\Omega_\oplus t_e/\la 1$,
the ground-based observations do not yield information about whether the lens
passed on the satellite side or the opposite side of the Earth.  
Thus, equation (\ref{eqn:satbeta}) contains essentially no information about
$\Delta u_x$.  It is not immediately obvious,
but I show in \S\ 3.2 that equation (\ref{eqn:satgamma}) also has relatively 
information about $\Delta u_x$.  Hence, in agreement with our naive
expectation, $\Delta u_x$ is mainly determined by measuring $\Delta t_0$.
However, from equation (\ref{eqn:sattnought}) we see that uncertainties
in the measurement of $\gamma_\oplus$ can propagate into the $\Delta u_x$
determination.

\subsection{Ground-based Observations}

	I have previously discussed in some detail the problem of early
identification and intensive monitoring of microlensing events toward the LMC
(Gould 1998b).  As in that paper, 
I evaluate the covariance matrix $c_{i j}$ of the
six parameters $a_i(=t_0$, $\beta$, $t_e$, $\gamma$, $F_0$, $B$) 
specified in equations (\ref{eqn:adef}), (\ref{eqn:foft}), and
(\ref{eqn:xpoft}), by considering a series of measurements at times $t_k$,
and with errors $\sigma_k$,
\begin{equation}
c = b^{-1},\qquad b_{i j} = \sum_k \sigma_k^{-2}
{\partial F(t_k)\over\partial a_i}\,{\partial F(t_k)\over\partial a_j}.
\label{eqn:bijdef}
\end{equation}
After taking the derivatives $\partial F(t_k)/\partial a_i$, I evaluate them
assuming $B=\gamma=0$.  I assume that the errors are photon limited, i.e.,
$\sigma_k = \sigma_0 F_0 [A(t_k)]^{1/2}$.  
(This assumption differs somewhat from Gould 1998b.)\ \
I assume that these intensive observations are triggered when
the event enters the Einstein ring ($u_{\rm init}=1$, $A(u_{\rm init})=1.34$) 
and end at
$t = t_0 + 1.5 t_e$, and that they are carried on uniformly at a rate
$N$ per day in the interval.  I then find an uncertainty in the determination 
of all parameters.  In particular, for the key parameters, $t_0$, $\gamma$,
and $\beta$.
I find
\begin{equation}
{\sigma_{t_0}\over t_e} = {\sigma_0\over (N t_e/\rm day)^{1/2}}R(\beta),
\label{eqn:sigmatnought}
\end{equation}
and
\begin{equation}
\sigma_\gamma = {\sigma_0\over (N t_e/\rm day)^{1/2}}S(\beta),\qquad
\sigma_\beta = {\sigma_0\over (N t_e/\rm day)^{1/2}}T(\beta),
\label{eqn:sigmagamma}
\end{equation}
where $R(\beta)$, $S(\beta)$, $T(\beta)$ are shown in Figure 
\ref{fig:one}.  

	These results shed light on three questions regarding equations
(\ref{eqn:deltauy}),
(\ref{eqn:sattnought}), and (\ref{eqn:satgamma}).  First, the ratio of the 
contributions of $\sigma_\gamma$ and $\sigma_{t_0}$ 
to the error in the determination of 
$\Delta u_x$ from equation (\ref{eqn:sattnought}) is
\begin{equation}
2(\Omega_\oplus t_e)^{-2}\sin^2\theta
{\sigma_\gamma\over \sigma_{t_0}/t_e}\simeq 0.4\biggl({d\over 0.2 \au}\biggr)^2
\biggl({t_e\over 40\,\day}\biggr)^{-2}\,{S(\beta)\over 10 R(\beta)}.
\label{eqn:comperr}
\end{equation}
Thus, for typical parameters, the two sources of error are comparable.
See Figure \ref{fig:one}.

	Next, I note that the error in $\Delta u_x$ induced by the
``$\gamma_\oplus$'' term in equation (\ref{eqn:satgamma}) is larger than
the error induced by the
``$\Delta t_0$'' term in equation (\ref{eqn:sattnought}) by a factor
$\sim S(\beta)/R(\beta)\sim 10$. 
Hence, for typical parameters, $\Delta u_x$ is constrained
primarily by equation (\ref{eqn:sattnought}).  This result has the important
consequence that the {\it SIRTF} observations should be optimized to
constrain $t_{0,S}$ rather than $\gamma_S$ or any other parameters 
(see \S 3.3).

	Finally, the error in $\Delta u_y$ is larger by a factor
$\sim 2(d/\au)^{-1}$ than the error induced by
the $\gamma_\oplus$ term in equation (\ref{eqn:sattnought}).  By 
equation (\ref{eqn:comperr}) the latter is comparable to the total error in 
$\Delta u_x$.  Hence, the (purely ground-based) error in $\Delta u_y$ is
substantially 
larger than the ground-based contribution to $\Delta u_x$.  That is,
the limit imposed by the ground-based observations on the overall precision
of the measurement is set by the ground-based measurement of $\Delta u_y$,
i.e., $\gamma$.  From equations (\ref{eqn:gammadef}), (\ref{eqn:deltauy}), 
and (\ref{eqn:sigmagamma}),
the error in $\Delta u_y$ expressed as a fraction of $\Delta u$ is
\begin{equation}
{\sigma_{\Delta u_y}\over \Delta u} = {\sigma_\gamma\over \gamma\sec\phi}
= 0.17\,N^{-1/2}\,{\sigma_0\over 0.01}\,{\tilde v\over 275\,\kms}\,
\biggl({t_e\over 40\,\day}\biggr)^{-3/2}\,{S(\beta)\over 8}
\label{eqn:siguyeval}
\end{equation}
Thus, for a robust detection of parallax for a halo event with typical
parameters requires $\sigma_0=1\%$ photometry, on average once per day.

\subsection{{\it SIRTF} Observations}

	There are a number of constraints that affect {\it SIRTF} observations
that do not affect ground-based observations, the most important of which is
scheduling.  While the operational plan for {\it SIRTF} is not finalized,
it is expected that the IRAC camera (used for $L$ band photometry) will 
share time equally with two other instruments, rotating for example one week
on, two weeks off.  Thus, the type of schedule envisaged for the 
ground-based observations (exposures once or several times per day) are out
of the question for {\it SIRTF}.  Instead, observations must be concentrated
in a few critical periods, each less than a week and separated by several
weeks.  This appears at first to pose major difficulties, but in fact such
constraints are perfectly compatible with an optimal observing plan.

	To devise an optimal strategy, first recall that the flux is a function
of time and six parameters, $F(t;a_i)$.  The derivatives of this flux with
respect to the parameters, $\partial F/\partial a_i$, are even in $(t-t_0)$
for four  parameters $(a_i = \beta,t_e,F_0,B)$ and odd for the other two
$(a_i=t_0,\gamma)$.  Hence, any set of observations that was symmetric in
$(t-t_0)$ would automatically decouple the errors in $(\beta,t_e,F_0,B)$
from those in $(t_0,\gamma)$.  This would be good because, from the discussion
in \S\ 3.1 and \S\ 3.2, $t_0$ and $\gamma$ provide essentially all of the 
useful
information.  I focus first on the simplest symmetric case, two observations
placed at $t_\pm=t_0 \pm \tau t_e$, where $\tau$ is a parameter.  In fact,
a third observation is required to establish the baseline, $F_0+B$.  I will
return to this point below, but for the moment I assume that the baseline
is known with perfect precision.

	Now, if it were literally true that there was {\it no} information
about $\beta$, $t_e$, and $F_0-B$, then of course a light curve fit to only two
points would be completely unconstrained.  However, from the ground-based
data and equations (\ref{eqn:satbeta}) and (\ref{eqn:satte}), $\beta_S$ and
$t_{e,S}$ are reasonably constrained.  I will discuss this in more detail
below.  
Here I focus on what specifically can be learned about $t_{0,S}$, and
$\gamma_S$ from this symmetric pair of observations.  The $(2\times 2)$
inverse covariance matrix associated with $t_0$ (normalized to $t_e$) and
$\gamma$ is (see eqs.\ \ref{eqn:adef}, \ref{eqn:foft}, and \ref{eqn:bijdef})
\begin{equation}
b_{i j}(t_0/t_e,\gamma) = {64\over u^5(u^2+4)^{5/2}(u^2+2)\sigma_0^2}
\left(\matrix{2\tau^2 & -\tau^4\cr -\tau^4 & \tau^6/2\cr}\right),
\label{eqn:redcovmat}
\end{equation}
where $u^2 = \tau^2 + \beta^2$.  This matrix is of course degenerate
between $t_0$ and $\gamma$.  I assume for the moment that the information
in equation ({\ref{eqn:satgamma}) is sufficient to break this degeneracy.
(I check this assumption below.)\ \ The error in $t_0/t_e$ from these two
measurements is then $\sigma_{t_0} \simeq [b(1,1)]^{-1/2}$.  This error is
minimized approximately at $\tau \sim (2/3)^{1/2}\beta$ at which point
\begin{equation}
{\sigma_{t_0} \over t_e} \sim \biggl({25\over 12}\biggr)^{1/2}\beta\sigma_*,
\quad(\sigma_*= ({5/3})^{1/4}\beta^{1/2}\sigma_0),
\qquad ({\rm at}\ \tau = (2/3)^{1/2}\beta)
\label{eqn:sigtnoughttwo}
\end{equation}
where $\sigma_*$ is the approximate
fractional flux error of the two observations, and where I have made the
evaluations using the approximation $A(u)\sim u^{-1}$.
Thus, for typical values, $\beta\sim 0.4$ and $t_e\sim 40\,$days, 
the two observations are separated by $\sim 26\,$days, and the 
error in $t_0$ is $\sigma_{t_0}/t_e\sim 0.6\sigma_*$.

\subsection{Correlations}

	To arrive at this estimate, I have argued or assumed that one can
ignore the numerous correlations among the 12 observable parameters (six from 
the Earth and six from the satellite).  I test these assumptions by simulating
a fit based on the observations as outlined in \S\ 3.2 and \S\ 3.2.  In order
to do so, I must choose a relative scale of errors for the Earth and satellite
observations.  The observations should 
be designed so that the errors in $\Delta u_x$ and $\Delta u_y$ are roughly 
comparable.  On the assumption that $[b(1,1)]^{-1/2}$ in equation 
(\ref{eqn:sigtnoughttwo}) gives a good estimate of the error in $t_0$ and
that this error dominates the error in $\Delta u_x$ (see eq.\
\ref{eqn:sattnought}), I initially assume that 
$(25/12)^{1/2}\beta\sigma_* = (d/\au)(\Omega_\oplus t_e)^{-2}\sigma_\gamma$ 
(see eq.\ \ref{eqn:deltauy}).  For definiteness, I initially
assume $d=0.2\,\au$ and 
$t_{e,\oplus} = 40\,$days.

	I fit for a total of 13 parameters including 12 observables
($a_i$, $i=1 \ldots 6$ 
for $t_0$, $\beta$, $t_e$, $F_0$, $B$, and $\gamma$ as seen 
from the Earth, and $a_i$, $i=7 \ldots 12$ for same parameters as seen from the
satellite) 
plus $a_{13} = \Delta u_x$ which is a derived parameter.  The inverse 
covariance matrix
$b_{i j}$ is given by the sum of four types of terms.  First, there are terms
of the form given by equation (\ref{eqn:bijdef}) for Earth-based observations
which affect $b_{i j}$, $i,j=1\dots 6$.  Second, there are terms of the
same form for satellite-based observations which affect 
$b_{i j}$, $i,j=7\dots 12$.  Third there are contributions to $b_{i j}$ from
the constraints (\ref{eqn:sattnought}), (\ref{eqn:satte}), and 
(\ref{eqn:satgamma}).  Each of these can be written in the form
$\sum_i \alpha_i a_i = 0$.  For example, for equation  (\ref{eqn:satgamma}),
$\alpha_{12} = 1$, $\alpha_{13}=-(\Omega_\oplus t_{e,\oplus})^{-2}\cos\theta$,
$\alpha_6 = -\cos 2\theta$, and $\alpha_i=0$ for all other $i$.  
These constraints
lead to contributions to $b_{i j}$ of the form $\alpha_i\alpha_j/Q^2$ where
$Q$ is an arbitrarily small number.  Finally, equation (\ref{eqn:satbeta})
leads to a similar constraint, except that there is a discrete uncertainty
in the sign of $\Delta \beta$.  Hence the constraint is less certain and
so has the form $\alpha_i\alpha_j/(\Delta\beta)^2$.  For definiteness, I 
choose $\Delta\beta= 5\sigma_{\Delta u_y}$ on the assumption that the
observations have been structured to detect $\Delta \beta$ at 
the $5\,\sigma$ level.

	I then find for a pair of observations that are exactly
symmetric about $t_{0,S}$, that for $\beta= 0.1 \ldots 0.5$, the errors are
higher than my naive expectations by factors $f= 1.03$ 1.07, 1.14, 1.24, and
1.36.  If the various assumptions that I have made in the analytic derivation
had all held exactly, these ratios would all be unity.  The deviations from
unity are partly accounted for by the fact that to derive equation
(\ref{eqn:sigtnoughttwo}), I evaluated equation (\ref{eqn:redcovmat}) assuming
$(u^2+4)^{5/2}(u^2+2)/64=1$, whereas it is actually slightly higher and
increases with increasing $\beta$.  Another factor is that in equation
(\ref{eqn:redcovmat}), $[b(2,2)/b(1,1)]^{1/2}=\beta^2/3$.  Thus, as $\beta$
increases the role of the $b(2,2)$ ($\gamma$) term is relatively
less well constrained by equation
(\ref{eqn:satgamma}).  In brief, the estimate (\ref{eqn:sigtnoughttwo})
for the required photometric precision is basically accurate but is slightly
too optimistic for $\beta\ga 0.3$.

	However, the assumption that the pair of observations is {\it exactly}
symmetric about $t_{0,S}$ is too idealized.  Even if one had perfect freedom
to schedule the observations, and even if one knew $t_{0,\oplus}$, 
$t_{e,\oplus}$, and $\gamma_\oplus$ exactly
from the ground-based observations, according to equation 
(\ref{eqn:sattnought}) there would still be an uncertainty 
$\Delta u_x t_{e,\oplus}\cos\theta$ in the predicted value of $\Delta t_0$
(and so of $t_{0,S}$),
where $\Delta u_x$ is an unknown quantity still to be measured.  For typical
parameters, $\Delta t_0\sim d/ \tilde v\sim {\cal O}(1\,\day)$.  In
addition, at the time that the second observation is planned, there will
be some measurement uncertainties in $t_{0,\oplus}$, 
$t_{e,\oplus}$, and $\gamma_\oplus$.  This will lead to an additional
uncertainty in the prediction of $t_{0,S}$, although this uncertainty will
probably be less than 1 day.  The biggest potential problem is that it may
not be possible to schedule the second set of observations exactly when one
would like.

	Regardless of the reason, if the observations are not symmetric about
$t_{0,S}$, then the errors in $\beta$, $t_e$, and $F_0-B$ will not decouple
from those in $t_0$, and $\gamma$.  These three quantities are relatively
poorly determined, so the degradation of the precision could
be significant if the asymmetry of the observations about the peak
is sufficiently large.  To quantify
this effect, I imagine that the two observations take place at
$t_\pm = t_0 + \delta t \pm (2/3)^{1/2}\beta t_e$.  That is, their midpoint
is displaced from $t_0$ by a time $\delta t$, but the separation between the
observations is the same as in the optimal case (eq.\ \ref{eqn:sigtnoughttwo}).
(The effect of using a different separation can easily be judged from the (1,1)
component eq.\ \ref{eqn:redcovmat}).\ \ 
Since the errors are a minimum for $\delta t=0$,
we expect that they will be quadratic in $\delta t$.

	Figure \ref{fig:two} shows the ratio, $f$, of the true error
in $\Delta u_x$ to the naive error
as a function of $\delta t$ for
$\beta=0.2,$ 0.3, 0.4, and 0.5.  For definiteness, I have as before assumed
$d=0.2\,\au$ and $t_{e}=40\,$days.  The error in $\Delta u_x$ becomes
seriously degraded if $\delta t$ is more than about two days.  The effect is
worse for low $\beta$ because the effective time scale $t_{\rm eff}=\beta t_e$
is shorter, so the asymmetry of the observations is more severe for fixed
$\delta t$.  This result emphasizes the importance of scheduling the
observations to be as symmetric as possible.  However, it also shows that
the inevitable $\sim 1\,$day errors in estimating $t_{0,S}$ will not seriously
affect the precision.

	If the satellite separation is reduced to $d=0.1\,\au$, but
$t_e$ remains at 40 days, then the results shown in Figure \ref{fig:two}
remain qualitatively the same.  On the other hand, if $d$ remains at $0.2\,\au$
while $t_e$ is reduced to 20 days, then all the curves rise about twice
as rapidly.  That is, the figure remains approximately accurate for all 
relevant values of $d$ and $t_e$ provided that the
abscissa is labeled ``$(\delta t/t_e) \times (40\,$days)''.

\subsection{Constraining the Baseline}

	The calculations of the previous section were somewhat idealized
in that they assumed that the baseline ($F_0+B$) is known exactly.  For
satellite observations such exaggerated precision would come at a very
high cost.  In fact, observations of the baseline need not be very intensive.
This can be understood as follows.  Let the fluxes measured on opposite
sides of the peak be $F_1$ and $F_2$, and suppose that the measurements are
nearly symmetric so that $\Delta F \equiv F_1-F_2\ll F_*\equiv (F_1+F_2)/2$.
From the ground-based measurements combined with equations 
(\ref{eqn:sattnought})-(\ref{eqn:satgamma}) (plus the fact that $|\Delta \bu|
\ll 1)$, one knows $t_{0,S}$, $\beta_S$, $t_{e,S}$, and $\gamma_S$ to within
a few percent, and hence one also knows $A$ at the time of the observations
with similar precision.  The quantity that gives information about 
$\Delta t_0$ is $\delta A/A = \delta F/(A F_0)$.  That is, uncertainty in
the estimate of $F_0$ will degrade the precision of $\Delta t_0$ only if
its fractional error is of the same order or larger than the fractional
error in $\Delta t_0$.  Since the latter is not likely to be much lower
than $\sim 20\%$, only a relatively crude measurement of $F_0$ is necessary.
The near-peak measurements give $A F_0 + B$ fairly precisely, and the
baseline gives $F_0+B$.  The error in the difference, $(A-1)F_0$, will thus
be of the same order as the error in the baseline.  Since $A$ is known
relatively well, $F_0$ will be determined with similar fractional accuracy
as the baseline.  Thus, the baseline measurement can be an order of magnitude
or more less precise than the peak measurements.

	I find numerically that if the baseline exposure time is equal
to the exposure time for each of the near-peak observations, then the
precision of the determination of $\Delta u_x$ is affected by 1\% or less.
Even if the exposure time is reduced by a factor 10, the precision of
$\Delta u_x$ is degraded by 10\% or less.  Thus, the total required
observation time is well approximated by the time required for the two
near-peak observations.

\section{Practical Considerations}

	Here I consider the problems of timely event recognition, 
signal-to-noise ratio (S/N), image analysis, telescope time requirements, 
and backgrounds.

\subsection{Event Recognition}

	The event must be recognized sufficiently early for two reasons.
First, the precision of the ground-based measurement of $\gamma_\oplus$ 
(and so $\Delta u_y$) depends critically on when the intensive follow-up
observations begin.  Second, the characteristics of the event ($t_0$, $\beta$,
$t_e$, and $F_0$) must be sufficiently well understood from the initial
ground-based follow-up observations to plan the satellite observations.
These latter are, in their nature, target-of-opportunity observations and
so will inevitably require some rescheduling.  Moreover, given the rotation
of instruments, there is only a 1/3 probability that the IRAC camera will
be scheduled for use at the optimal time for the first exposure and another,
independent 1/3 probability that it will be scheduled for use at the
optimal time for the second exposure.  Thus, a complicated decision process
will be necessary to balance the requirements of these observations with other
aspects of the {\it SIRTF} mission, and it is important that sufficient
information about the event be available to make a rational decision
on a timely basis.

	Early recognition presents a significant challenge for the proposed
observations.  At present, the MACHO collaboration alerts on events when
they surpass magnification $A=1.6$ ($u_{\rm init}=0.75$) rather than 
$A=1.34$ ($u_{\rm init}=1$) (A.\ Becker 1998, private communication).
Figure \ref{fig:three} compares the S/N function $S(\beta)$ 
(see eqs.\ \ref{eqn:sigmagamma} and \ref{eqn:siguyeval}) for these two
values of $u_{\rm init}$.  It is clear from this figure
that the S/N is severely degraded for
$\beta \ga 0.45$ at $u_{\rm init}=0.75$.
A closely related
problem is that the optimal time for the initial satellite observation
is at $u=(5/3)^{1/2}\beta=1.29\beta$ (see eq.\ \ref{eqn:sigtnoughttwo}).
Thus, for $\beta=0.5$, the optimal time for the first satellite observation
is at $u=0.65$ which is only a time $0.15\,t_e\sim 6\,$days after the
present-day initial alert.

	It is therefore important, though not absolutely critical, to
alert on events earlier than the present standard.  One approach would be
to initiate aggressive follow-up observations at a lower threshold, and
to weed out the false alerts through these intensive observations.  This would
probably require a dedicated or nearly-dedicated 1 m telescope.  Another
approach would be to obtain higher S/N during the
initial microlensing search observations.  C.\ Stubbs (1998 private 
communication) and collaborators are trying to organize a search with a
2.5 m telescope, $1''$ seeing, and a 1 deg${}^2$ camera, which would 
represent a factor 7 improvement in S/N relative to the present MACHO setup.
It might well then be possible to alert on events at lower magnification.
However, even if neither of these more aggressive programs are implemented,
one could still carry out parallax measurements by restricting events to
those with $\beta\la 0.45$.

\subsection{Image Analysis}

	I assume that the observations will be analyzed using 
image-differencing techniques which have been pioneered by Tomaney \& Crotts
(1996) and Ansari et al.\ (1997) to find microlensing events of unresolved
stars in M31.  Melchior et al.\ (1998), Tomaney (1998), and Alard \& Lupton 
(1998) have further refined these techniques for application to 
photometry of {\it resolved} (or partially resolved) sources such as those
that are routinely monitored in the LMC, SMC, and Galactic bulge.  A version
of the Alard-Lupton algorithm is now used by the EROS collaboration
to make precise light-curve measurements for events found using their normal
(more standard) photometry (Afonso et al.\ 1998).

	The application of these techniques to the {\it SIRTF} IRAC camera
should be relatively straight forward, but not completely trivial.  
The pixel size of this camera
is $1.\hskip-2pt'' 2$ while the point spread function (PSF) is about $2''$.
This is somewhat larger than the diffraction limit at $3.6\,\mu$m 
(with a 0.85 m telescope) of $\sim 1''$.  However, the PSF will still be 
undersampled.  The two exposures that are to be differenced 
will be separated by $\sim 25\,$days and therefore rotated relative to one
another by $\sim 25^\circ$.  Hence, the source star will not fall on exactly
the same parts of the pixels in the two exposures.  
Variations in sensitivity across a single
pixel would therefore make it impossible to obtain the required photometric 
precision (see \S\ 4.3) by point-and-stare observations.  Hence, it will be
necessary to dither the telescope by fractions of a pixel many times in
order to enlarge the effective PSF and so smooth over the pixel-sensitivity 
variations.  

	The normal practice in image differencing is to form a ``template
image'' by combining several images at baseline, and then to subtract this
from the event images.  For the present case the procedure is quite different.
The main information comes from directly differencing the two images
taken symmetrically about the peak of the event.  The shorter baseline
exposure is {\it not} used as a template.  Rather, the difference between
this image and the average of the two images near peak is used to extract
the less precise auxiliary information about $F_0$.  See \S\ 3.5.

	I take note of a minor technical point.  Image differencing
automatically removes the background sources $B$ from the analysis.  This
means that a standard microlensing light curve is fit to four parameters
rather than five.  That is, one fits to
\begin{equation}
\tilde F(t;t_0,\beta,t_e,F_0) = F_0[A(t:t_0,\beta,t_e)-1],
\label{eqn:pixel}
\end{equation} 
instead of
equation (\ref{eqn:foft}). For this reason, it is sometimes mistakenly believed
that the fit is less affected by the degeneracies associated with background
light.  Recall from \S\ 2, that removal of these degeneracies is the central
problem addressed by this paper.  Why is equation (\ref{eqn:pixel}) no more
constraining than equation (\ref{eqn:foft})?  The latter can be formally
rewritten 
\begin{equation}
F(t;t_0,\beta,t_e,F_0,\tilde B) = F_0[A(t:t_0,\beta,t_e)-1] +
\tilde B,
\label{eqn:pixeltwo}
\end{equation} 
where $\tilde B\equiv F_0+B$ is the baseline flux.  In normal
microlensing observations, $\tilde B$ is extremely well determined because
there are a large number of observations at baseline.  Hence, it is effectively
decoupled from the other parameters in equation (\ref{eqn:pixeltwo}).  On
the other hand, $\tilde F$ in equation (\ref{eqn:pixel}) refers to the
flux {\it above baseline}.  That is, equation (\ref{eqn:pixel}) 
implicitly assumes that the
baseline is well observed and that an essentially perfect template has been
formed from these observations.  Thus, in reality, equation
(\ref{eqn:pixel}) has the same information content as
equation (\ref{eqn:foft}) or equation (\ref{eqn:pixeltwo}) 

\subsection{Telescope Time}

	I have already discussed at some length the requirements for the
ground-based observations to measure $\gamma_\oplus$ and thus $\Delta u_y$
(Gould 1998b).  The
basic result is summarized by equation (\ref{eqn:siguyeval}).  I focus
here on the requirements for {\it SIRTF} observations.  

	In \S\ 3.4, I 
introduced the quantity $\sigma_*$, the fractional photometry error for 
symmetric pair of observations near the peak.  Recall that the error in
$\Delta u_x$ is given by $\sigma_{\Delta u_x}=
(25/12)^{1/2}f\beta \sigma_*$, where $f$ is the correction 
factor shown in Figure \ref{fig:two}.  For reasonably symmetric pairs of
observations, $1\leq f\leq 1.36$
The companion to equation 
(\ref{eqn:siguyeval}) is then
\begin{equation}
{\sigma_{\Delta u_x}\over \Delta u} = \biggl({25\over 12}\biggr)^{1/2}
f\beta\sigma_*{\tilde v t_e\over d}
= 0.18f\,{\sigma_*\over 0.01}\,{\beta\over 0.4}\,{\tilde v\over 275\,\kms}\,
{t_e\over 40\,\day}\,\biggl({d\over 0.2 \au}\biggr)^{-1}.
\label{eqn:siguxeval}
\end{equation}
Thus, for a robust detection of parallax for a halo event with typical
parameters requires $\sigma_*=1\%$ photometry for two each of two
observations near the peak, plus a shorter exposure of the baseline (see
\S\ 3.5).  This is to be compared with the ground-based requirement of
1\% photometry, once per day for a period $\sim 2.5\, t_e$.

The IRAC detector records $0.7\,e^-\,{\rm s}^{-1}\,\mu{\rm Jy}^{-1}$,
and the sky plus dark current is expected to be 
$3\,e^-\,{\rm s}^{-1}\,\rm pixel^{-1}$ (J.\ Hora 1998 private communication).
The pixel size is $1.\hskip-2pt '' 2$.  
The PSF is expected to be $2''$.
However, I assume that the effective size of the PSF is
increased to $3''$ by dithering (see \S\ 4.2).  
These figures imply that the background is $\sim 45\,e^-\,\rm s^{-1}$.

	For the great majority of events detected to date, the source star
is $V\geq 20$.  For illustration, I consider two $V=20$ stars, a 
main-sequence star ($V-L=0.3$) and a clump giant ($V-L=2.5$).  
I assume that $1\mu\rm Jy$ corresponds to $L=21.1$.  For the
main sequence star, I find that the total exposure time required to
reach $\sigma_*=1\%$ precision is $T_{\rm exp} = 39(\beta^2 + 0.04\beta)\,$hr.
For the clump giant, the time required is
$T_{\rm exp} = 0.6(\beta^2 + 0.33\beta)\,$hr.
Thus, for main sequence stars with $\beta\la 0.4$, the total satellite time
is $\sim 2 T_{\rm exp}\la 12\,$hours, while for a red giant, the time required
is less than one hour.  Clearly, the latter is to be preferred.  To date,
unfortunately, only one LMC clump giant event has been published 
(Alcock et al.\ 1998a).

\subsection{Backgrounds}

	As I have previously discussed (Gould 1998b), an asymmetric light
curve can be the result of a binary lens or a binary source.  I made a
rough estimate that $\sim 20\%$ of events could be affected by such 
backgrounds which would lead to a spurious measurement of $\Delta u_x$.
There are no backgrounds that would mimic a shift in the peak time as seen
by the satellite relative to the Earth, assuming that the peaks were well
resolved.  Nevertheless, the method proposed here is to determine the peak from
a pair of symmetrically placed observations.  Thus, the same asymmetry that
produces a spurious detection of $\gamma_\oplus$, could in principle produce
a slight shift in $t_{0,S}$.  However, $\gamma$ is measured from the
``wings'' of the light curve, $(t-t_0)/t_e\sim\pm 1$, (Gould 1998b), while
the peak is determined from observations at 
$(t-t_0)/t_e\sim\pm (2/3)^{1/2}\beta$.  Since 
$\partial F/\partial\gamma\propto 0.5(t-t_0)^3$ while
$\partial F/\partial t_0 \propto 
(t-t_0)^1$, the effect of any background on the
$t_0$ determination will be $\sim \beta^2/3$ smaller than on the $\gamma$
determination.  That is, it will most likely be negligible.  Thus, it
would seem advisable to push the satellite observations so that 
$\sigma_{\Delta u_y} < \sigma_{\Delta u_x}$.  If the asymmetry detected
from the ground is truly due to parallax, one should also be able to
detect a time delay provided that 
the observations are sensitive enough.  Thus, the
satellite observations provide a partial check on the reality of the
ground-based detection of parallax.

\section{Other Lines of Sight}

	In this paper I have focused attention mainly on the LMC, partly
for simplicity and partly because I consider it to be the most interesting
line of sight scientifically.  However, there are two other lines of sight
for which one might want to obtain satellite parallaxes: the SMC and 
Galactic bulge.

	The scientific question regarding events detected toward the SMC is 
similar to that for LMC events: are the lenses in the halo or in the
Magellanic Clouds?  The major difference between the SMC and LMC is that
the LMC lies almost exactly at the ecliptic pole, whereas the SMC lies about
$25^\circ$ from the pole.  This difference in turn has two implications.
First, the equations describing parallax become more complicated.  See,
for example, Gaudi \& Gould (1997).  The ``parallax ellipse'' becomes more
flattened, making the parallax effects smaller and so more difficult to
measure.  However, since the
flattening is only by a factor $\cos 25^\circ\sim 0.9$, the effect is
quite minor and can be ignored for our purposes.  Second, {\it SIRTF} cannot
observe the SMC for the full year as it can the LMC.  The telescope can
only point between $80^\circ$ and $120^\circ$ from the Sun.  Hence the SMC
is only observable for $\sim 65\%$ of the year.  This is not a major
problem, but combined with the fact that the SMC is smaller than the LMC
(and so has fewer source stars to monitor) it does mean that it will provide
fewer opportunities for parallax measurements.

	The bulge is qualitatively different.  First it lies very close to
the ecliptic which implies that it can be observed only for two 40 day
intervals during the year.  Second, the parallax ellipse is highly flattened.
For example, Baade's Window lies only $6^\circ$ from the ecliptic
so the ellipse is flattened by a factor of 10.  This flattening will 
reduce the size of the parallax asymmetry by a factor of 10 in the summer
and winter, and will reduce the time delay between the Earth and satellite
by a factor 10 in the spring and fall (Gaudi \& Gould 1997).  Since the
bulge is observable from {\it SIRTF} only during the spring and fall,
it is the latter effect that is relevant.  Thus, equation
(\ref{eqn:siguyeval}) will be virtually unaffected, but equation
(\ref{eqn:siguxeval}) will be increased by a factor of 10.  This 
degradation of the S/N is partially
mitigated by the fact that bulge events tend to be shorter ($t_e\sim 10\,$days
rather than 40 days), but it is exacerbated by almost as large a factor
because most of the lenses 
are expected to be in the bulge implying that
$\tilde v= (\dos/\dls)v\sim 800\,\kms$ (rather than $275\,\kms$).  
Thus, the photometric precision required is increased by approximately a 
factor of 7.  From the standpoint of S/N, this is not a major problem.  The
bulge is about 6 times closer than the LMC, so clump giants are about 36
times brighter.  Hence, the exposure time is formally only
$T_{\rm exp} = 0.3\beta\,$hours.  However, whether systematics will compromise
photometry at the required 0.14\% level remains an open question.

	From a scientific standpoint, bulge parallaxes can address two 
principal questions.  
First, where are the lenses?  The conventional wisdom is that
most are in bulge.  However, the same conventional wisdom predicts a much
lower optical depth and many fewer short events than are actually observed.
One would like some experimental confirmation of this wisdom.  Second, what
is the mass spectrum of the lenses?  Again, the conventional wisdom is that
the lensing events are due to normal stars in the bulge (and secondarily the
disk) along the line of sight.  However, the observed mass spectrum of bulge
stars (Holtzman et al.\ 1998) does not seem to be able to explain the
observed distribution of time scales (Han \& Gould 1996).  Han \& Gould (1995)
showed that parallax measurements could help constrain both the location
and the mass spectrum of the lenses.

	Even if the time differences $\Delta t_0$ are initially too small
to measure, the situation will gradually improve with time because the
Earth-satellite distance $d$ will gradually grow.  This tends to increase
$\Delta t_0$ in two distinct ways.  First, of course, as the satellite
gets further from the Earth, it takes longer for a lens to move from one
to the other (see eq.\ \ref{eqn:siguxeval}).  In addition, the
satellite is constrained to observe the bulge not when it is spring or fall
on Earth, but when the Sun is near the vernal or autumnal equinox as seen
from the satellite.  As the satellite moves farther from the Earth, the
Earth-satellite separation vector becomes less closely aligned with the
direction of the bulge during these critical times that the bulge is
observable.  Unfortunately, these same changes also make it more difficult
to measure the parallax asymmetry from the ground.  Nevertheless, if it
initially proves too difficult to measure bulge parallaxes, the situation
should be reviewed periodically in light of ongoing experience.

\section{Conclusions}

	The method previously developed for measuring microlens parallaxes
by directly comparing Earth-based and satellite-based photometry will not
work for {\it SIRTF}.  The old method requires that both sets of 
observations be done in the same band in order to remove degeneracies in
the parallax solution.  The {\it SIRTF} $L$ band detector response cannot
be mimicked from the ground because of atmospheric absorption and high
background.

	However, by combining ground-based measurements of the 
``parallax asymmetry''
$\gamma$ of a lensing event (due to the Earth's acceleration) with the observed
difference $\Delta t_0$
in the peak times of the event as seen from the Earth and {\it SIRTF}, it
is possible to measure the parallaxes of microlensing events seen toward
the LMC.  The parallax yields the projected velocity of the lens,
$\tilde \bv = (\dos/\dls)\bv$ and so would reveal whether the lenses are
in the Galactic halo or in the LMC itself.  

	The ground-based measurement
requires $\sim 1\%$ photometry about once per day for about 2.5 Einstein
crossing times, i.e., 
$\sim 100\,$days.  The space-based measurement requires two observations
each with about 1\% precision, plus one additional lower-quality
measurement.  The two 1\% measurements should be spaced approximately
symmetrically about the peak of the event, while the lower-quality measurement
is needed to constrain the baseline.  For a main-sequence star ($V=20$,
$V-L=0.3$), the total satellite time required is 
$\sim 12\,{\rm hours}\, (\beta/0.4)(d/0.2\,\au)^{-1}$, where $\beta$ is the
impact parameter and $d$ is the Earth-satellite distance.  For a clump giant
($V=20$, $V-L=2.5$), less than 1 hour is required.

	Events must be alerted in real time and an improvement in the
current magnification threshold $(A\sim 1.6)$ for the alerts would be
helpful but is not critical.  Photometry should be carried out using
image differencing.  Backgrounds due to
binary lenses and binary sources are minor but not completely negligible.

	Parallax measurements are also possible for SMC events.  The major
difference from the LMC is that the SMC can be observed for only 70\% of the
year.  Parallax measurements of Galactic bulge microlensing events are
substantially different.  Formally, the telescope time 
requirements are less severe
than toward the LMC primarily because the sources are substantially brighter.
However, because the bulge lies near the ecliptic, the size of the
parallax effect is substantially smaller than toward the LMC, and this may
mean that small systematic errors will compromise the measurements.

{\bf Acknowledgements}:  
I thank G.\ Fazio and M.\ Werner for numerous stimulating
discussions and for their general enthusiasm for using {\it SIRTF} to
obtain microlens parallaxes.  I thank B.S. Gaudi for a careful reading
of the manuscript.
This work was supported in part by grant AST 97-27520 from the NSF and in 
part by grant NAG5-3111 from NASA.


\clearpage

\begin{figure}
\caption[junk]{\label{fig:one}
Normalized errors, $R$ ({\it bold}), $S$ ({\it solid}), 
and $T$ ({\it dashed}), for the measurements of
$t_0$, $\beta$, and $\gamma$.  These are respectively 
the time of maximum, the impact parameter, and the asymmetry parameter.
Each is plotted as a function of the impact parameter, $\beta$.  The
actual error is given by, for example, 
$\sigma_\gamma = S\sigma_0/(N t_e/\rm day)^{1/2}$
where $\sigma_0$ is the factional flux error for an individual measurement,
and $(N t_e/\rm day)$ is the number of measurements per Einstein crossing
time.  These curves assume that light-curve measurements begin when the source
enters the Einstein ring ($u_i=1.0$) and end at $t=t_0 + 1.5 t_e$.
}
\end{figure}

\begin{figure}
\caption[junk]{\label{fig:two}
Ratio $f$ of the actual error in $\Delta u_x$ (taking full account of all
the covariances among the 13 parameters of the full fit) to the naive error
given by equation (\ref{eqn:sigtnoughttwo}) (together with the approximations
that the error in $\Delta u_x$ is equal to the error in $\Delta t_0/t_e$, and
that the latter is dominated by the error in $t_{0,S}/t_e$).  This ratio
is shown for various values of the impact parameter $\beta$ as a function
of $\delta t$.  Here, $\delta t$ is the time difference between 
the peak of the event and the midpoint 
of the two observations by the satellite.  If
the two observations are symmetric about the peak $(\delta t = 0)$, then
$1\leq f \leq 1.36$.  However, the errors can increase dramatically for
non-symmetric observations, particularly for small $\beta$.  Symmetric
observations yield more precise results 
because the uncertainties in $t_e$, $\beta$, and
$F_0-B$ (which are all relatively large) are then decoupled from the 
uncertainties in $t_0$.  The curves shown in the figure all assume
an Earth-satellite separation $d=0.2\,\au$ and Einstein crossing time
$t_e=40\,$days.  However, for other values
of $t_e\la \yr/2\pi$ and $d\ll\au$, the curves are qualitatively similar.
}
\end{figure}

\begin{figure}
\caption[junk]{\label{fig:three}
Normalized error $S$ for the asymmetry parameter $\gamma$ which is used
to determine $\Delta u_y$.  See equations (\ref{eqn:deltauy}), 
(\ref{eqn:sigmagamma}), and (\ref{eqn:siguyeval}).  The lower curve assumes
that intensive follow-up observations begin when the enters the Einstein
ring $(u_{\rm init} = 1)$ and is the same as solid curve in Figure 
\ref{fig:one}.  This curve is assumed in all calculations in the paper.
The upper curve $(u_{\rm init} = 0.75)$ reflects the present capability of the
MACHO collaboration alert program.  If this capability cannot be improved,
the errors would increase substantially for $\beta\ga 0.45$.
}
\end{figure}

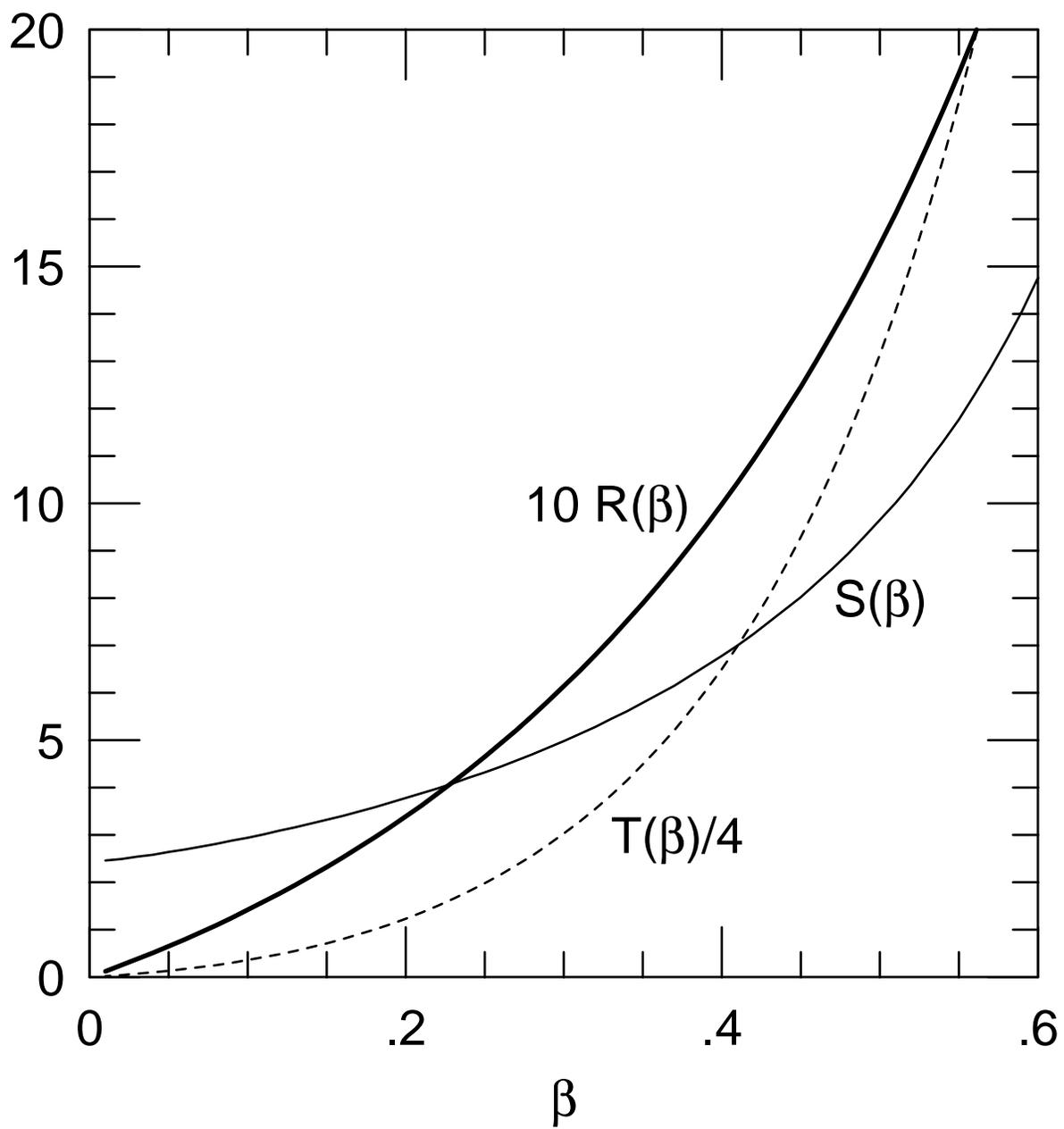
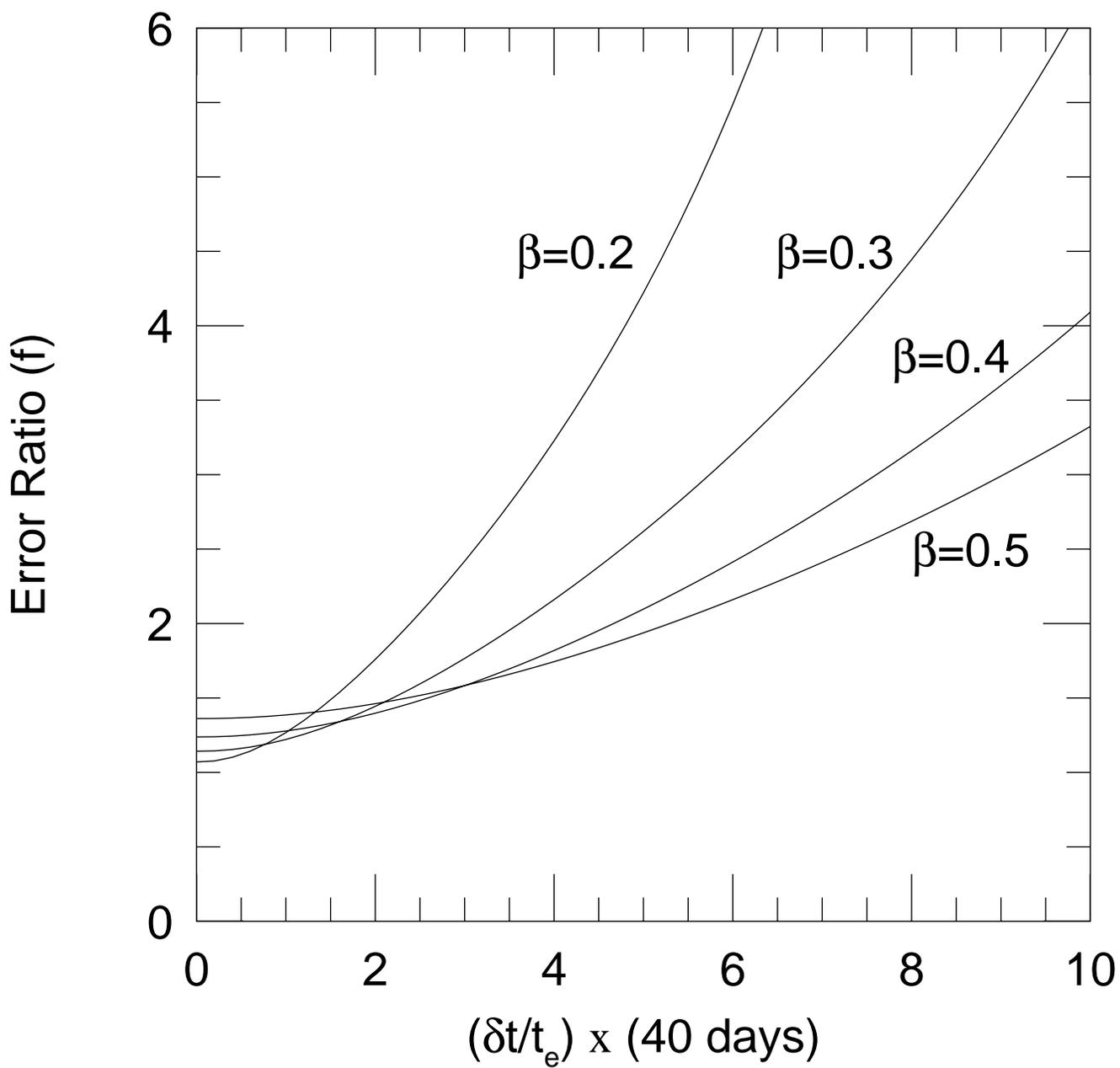
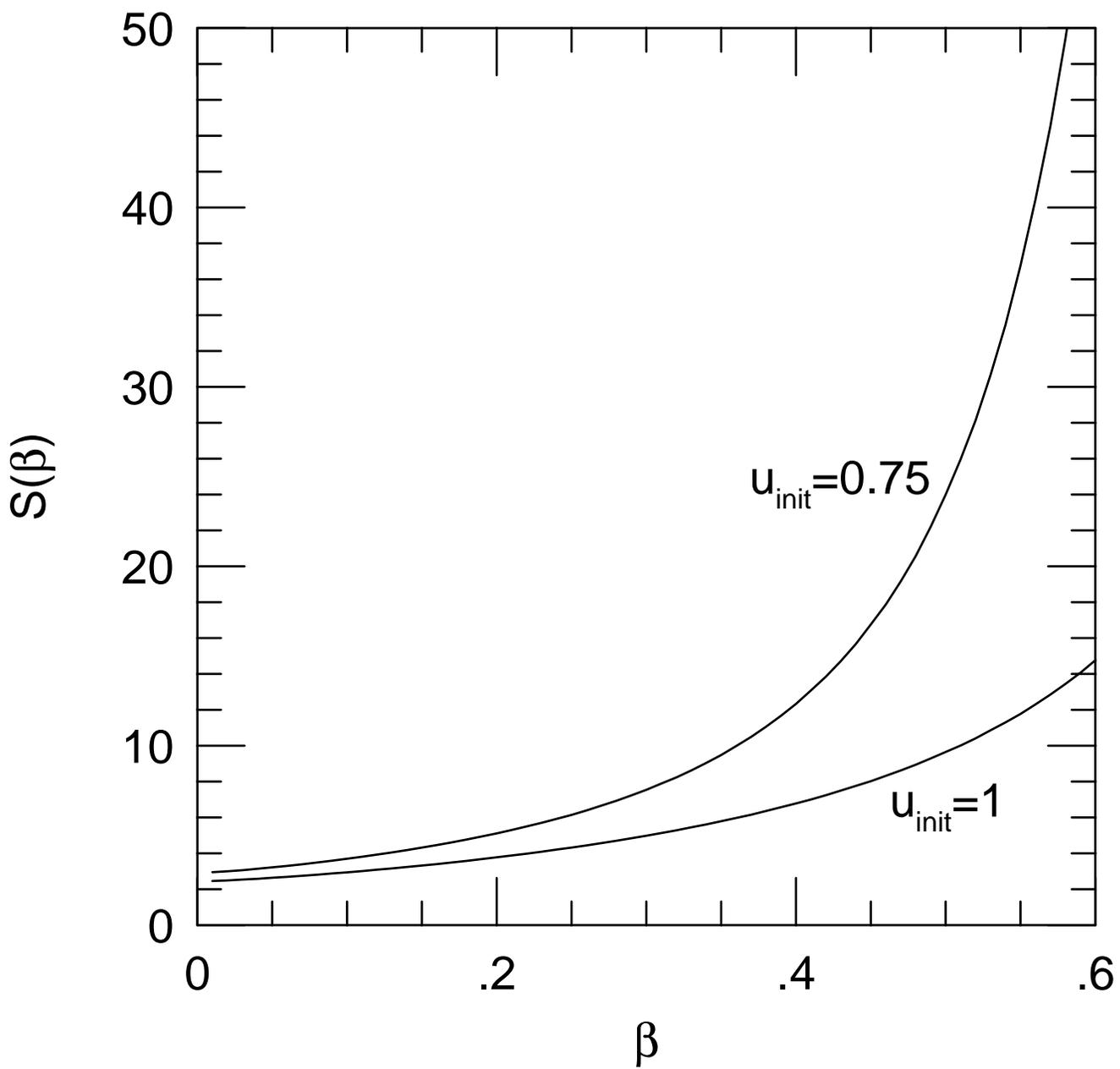
\end{document}